\documentstyle[aps,epsf,floats,twocolumn,pre]{revtex}
\begin{document}
\newcommand{\bgm}[1]{\mbox{\boldmath $#1$}}
\newcommand{\bgms}[1]{\mbox{{\scriptsize \boldmath $#1$}}}
\newcommand{\ul}[1]{\underline{#1}} \newcommand{\bgt}[1]{{\boldmath
    $#1$}}

\twocolumn[\hsize\textwidth\columnwidth\hsize\csname@twocolumnfalse%
\endcsname
\draft

\title{Occam factors and \\ model--independent Bayesian learning of
  continuous distributions}

\author{Ilya Nemenman$^{1,2}$ and William Bialek$^2$}
\address{$^1$Department of Physics, Princeton University,
  Princeton, New Jersey 08544\\
  $^2$NEC Research Institute, 4 Independence Way, Princeton, New
  Jersey 08540}

\date{\today}
\maketitle
\begin{abstract}
  Learning of a smooth but nonparametric probability density can be
  regularized using methods of Quantum Field Theory. We implement a
  field theoretic prior numerically, test its efficacy, and show that
  the data and the phase space factors arising from the integration
  over the model space determine the free parameter of the theory
  (``smoothness scale'') self--consistently. This persists even for
  distributions that are atypical in the prior and is a step towards a
  model--independent theory for learning continuous distributions.
  Finally, we point out that a wrong parameterization of a model
  family may sometimes be advantageous for small data sets.
\end{abstract}

\pacs{PACS}
]

\section{Introduction}
\label{intro}

One of the central problems in learning is to balance ``goodness of
fit'' criteria against the complexity of models.  An important
development in the Bayesian approach was thus the realization that
there does not need to be any extra penalty for model complexity: if
we compute the total probability that data are generated by a model,
there is a factor from the volume in parameter space---the ``Occam
factor''---that discriminates against models with more parameters
\cite{mackay,vijay} or, more specifically, against models which are
more complex in a precise information theoretic sense \cite{bnt}.
This works remarkably well for systems with a finite number of
parameters and creates a complexity ``razor'' (after ``Occam's
razor'') that is almost equivalent to the celebrated Minimal
Description Length (MDL) principle \cite{rissanen}.  In addition, if
the a priori distributions involved are strictly Gaussian, the ideas
have also been proven to apply to some infinite--dimensional
(nonparametric) problems \cite{gp}. It is not clear, however, what
happens if we leave the finite dimensional setting to consider
nonparametric problems which are not Gaussian, such as the estimation
of a smooth probability density.  A possible route to progress on the
nonparametric problem was opened by noticing \cite{bcs} that a
Bayesian prior for density estimation is equivalent to a quantum field
theory (QFT).  In particular, there are field theoretic methods for
computing the infinite dimensional analog of the Occam factor, at
least asymptotically for large numbers of examples.  These
observations have led to a number of papers
\cite{holy,periwalshort,periwallong,aida} exploring alternative
formulations and their implications for the speed of learning.  Here
we return to the original formulation of Ref.~\cite{bcs} and address
some of the questions left open by the previous work \cite{thesis}:
What is the result of balancing the infinite dimensional Occam factor
against the goodness of fit? Is the QFT inference optimal in using all
of the information relevant for learning \cite{bnt}?  What happens if
our learning problem is strongly atypical of the prior distribution?

The conclusions we finally make were not expected by us at the start
of the project, and they will probably be not intuitively obvious to
most of our readers either. Thus we chose to present this work in the
same way it had originally proceeded. First we develop a numerical
scheme for implementation of the learning algorithm of
Ref.~\cite{bcs}. Then we show some results of Monte--Carlo simulations
with this algorithm and notice some peculiar features that have not
been predicted by the previous literature.  Concurrently with the
simulations, we present a simple analytical argument that explains
these unexpected but extremely desirable features.

\section{Preliminaries}
\label{prelims}

Following Ref.~\cite{bcs}, if $N$ independent, identically
distributed samples $\{x_i\}, i=1\dots N,$ are observed, then the
probability that a particular density $Q(x)$ gave rise to these data
is given by
\begin{equation}
  P[Q(x)| \{x_i\}]
  = \frac{ {\mathcal P}[Q(x)] \prod_{i=1}^N Q(x_i)}{\int[dQ(x)] 
    {\mathcal P}[Q(x)] 
    \prod_{i=1}^N Q(x_i)}\,,
  \label{posterior}
\end{equation}
where ${\mathcal P}[Q(x)]$ encodes our a priori expectations of $Q$.
Specifying this prior on a space of functions defines a QFT, and the
optimal least square estimator is then the a posteriori Bayesian
average
\begin{equation}
Q_{\rm est}(x|\{x_i\})=\frac{\langle Q(x)Q(x_1)Q(x_2)\dots
  Q(x_N)\rangle^{(0)}}
{\langle Q(x_1)Q(x_2)\dots Q(x_N) \rangle^{(0)}}\,,
\label{solution}
\end{equation}
where $\langle\dots\rangle^{(0)}$ means averaging with respect to the
prior. Since $Q(x) \ge 0$, it is convenient to define an unconstrained
field $\phi(x)$, $Q(x)\equiv (1/\ell_0) \exp [-\phi(x)]$, where the
choice of the dimension setting constant $\ell_0$ must not influence
any final results. Other definitions are also possible \cite{holy},
but we think that most of our results do not depend on this choice.

Next we should select a prior that regularizes the infinite number of
degrees of freedom and allows learning.  We want the prior ${\cal
  P}[\phi]$ to make sense as a continuous theory, independent of
discretization of $x$ on small scales. Since it is not clear what a
renormalization procedure for a probability density would mean, we
also require that when we estimate the distribution $Q(x)$ the answer
must be everywhere finite.  These conditions imply that our field
theory must be ultraviolet (UV) convergent.  For $x$ in one dimension,
a minimal choice is
\begin{equation}
\label{prior}
{\cal P}[\phi(x)]= \frac{{\rm e}^{-\frac{\ell^{2\eta -1}}{2}\int
dx
\left(\frac{\partial^{\eta} \phi}{\partial x^{\eta}}
\right)^2}}{\cal Z}\, \delta \left[\frac{\int dx\, {\rm
e}^{- \phi(x)}}{l_0} -1 \right],
\end{equation}
where $\eta>1/2$, ${\cal Z}$ is the normalization constant, and the
$\delta$-function enforces normalization of $Q$. We refer to $\ell$
and $\eta$ as the {\em smoothness scale} and the {\em exponent},
respectively; they would be called {\em hyperparameters} in other
machine learning literature \cite{gp}.

In \cite{bcs} this theory was solved for large $N$ and $\eta=1$ using
the familiar WKB techniques. The saddle point (or the classical
solution) for the $\phi$ averaging in $\langle \prod_{i=1}^N Q(x_i)
\rangle^{(0)}$ was found to be given by
\begin{equation}
  \ell \partial^2_x \phi_{\rm cl}(x) + \frac{N}{\ell_0}
  {\rm e}^{-\phi_{\rm cl}(x)}
  = \sum_{j=1}^N \delta(x-x_j)\,,
  \label{stationary}
\end{equation}
and the fluctuation determinant around this saddle is
\begin{equation}
  R=\exp \left[-\frac{1}{2} 
    \sqrt{\frac{N}{\ell \ell_0}} \int dx \, 
    {\rm e}^{-\phi_{cl}(x)/2}\right]\,.
  \label{det}
\end{equation}
Then the correlation functions take a familiar form:
\begin{eqnarray}
  \langle \prod_{i=1}^N &&Q(x_i) \rangle^{(0)} \approx \frac{1}
  {\ell_0^N}\exp\left( -S_{\rm eff} [\phi_{\rm cl} (x) ; \{x_i\}]\right),
  \label{dist}
  \\
  S_{\rm eff}  &&=
\frac{\ell}{2}\int dx (\partial_x \phi_{\rm cl})^2  
+ \sum_{j=1}^N \phi_{\rm cl}(x_j) -\log R,
\label{action}
\end{eqnarray}

In Ref.~\cite{bcs} it was shown that, with such correlation functions,
Eq.~(\ref{solution}) is a ``proper'' solution to the learning problem:
it is nonsingular even at finite $N$, it converges to the target
distribution $P(x)$ that actually generates the data, and the variance
of fluctuations around the target, $\psi(x)\equiv -\log Q_{\rm est}(x)
- [-\log \ell_0 P(x)]$, falls off rather quickly as $\sim 1/\sqrt{\ell
  N P(x)}$.  It was also noted that the effective action
[Eq.~(\ref{action})] has acquired a term $-\log R$, which grows as
$\ell$ decreases. This is contrary to the data contribution,
$\sum_{j=1}^N \phi_{\rm cl}(x_j)$, which favors small $\ell$ and the
corresponding overfitting.  Thus the $-\log R$ term may be rightfully
called an infinite dimensional generalization of the Occam factors.
The authors speculated that, if the actual $\ell$ is unknown, one may
average over it and hope that, much as in Bayesian model selection
\cite{mackay,vijay}, the competition between the data and the
fluctuations will select the optimal smoothness scale $\ell^*$.
Finally, they suggested that this optimal scale might behave as
$\ell^* \sim N^{1/3}$.

Before we proceed on to the numerical implementation of the above
algorithm, a note is in order. At first glance the theory we study
seems to look almost exactly like a Gaussian Process \cite{gp}. This
impression is produced by a Gaussian form of the smoothness penalty in
Eq.~(\ref{prior}), and by the fluctuation determinant that plays
against the goodness of fit in the smoothness scale (model) selection.
However, both similarities are incomplete. The Gaussian penalty in the
prior is amended by the normalization constraint, which gives rise to
the exponential term in Eq.~(\ref{stationary}), and violates many
familiar results that hold for Gaussian Processes, the representer
theorem \cite{representer} being just one of them. In the
semi--classical limit of large $N$, Gaussianity is restored
approximately, but the classical solution is extremely non--trivial,
and the fluctuation determinant is only the leading term of the
Occam's razor, not the complete razor as it is for a Gaussian Process.
In addition, it depends on the data only through the classical
solution; this is remarkably different from the usual determinants
arising in the Gaussian Processes literature \cite{gp,holy}.

\section{The algorithm}

Numerical implementation of the theory is rather simple.  First, to
eliminate a possible infra--red singularity in Eq.~(\ref{det})
\cite{bnt,thesis}, we confine $x$ to a box $0\le x\le L$ with periodic
boundary conditions. The boundary value problem Eq.~(\ref{stationary})
is then solved by a standard ``relaxation'' (or Newton) method of
iterative improvements to a guessed solution \cite{numrec} (for the
target precision we always use $10^{-5}$). The independent variable $x
\in [0,1]$ is discretized in equal steps [$10^4$ for
Figs.~(\ref{correct_ex}--\ref{incorrect_diff}), and $10^5$ for
Figs.~(\ref{scale_select}, \ref{adaptive_D})].  We use an equally
spaced grid to ensure stability of the method, while small step sizes
are needed since the scale for variation of $\phi_{\rm cl}(x)$ is
\cite{bcs}
\begin{equation}
\delta x \sim \sqrt{\ell/ N P(x)}\,,
\label{varscale}
\end{equation}
which can be rather small for large $N$ or small $\ell$. 

Since the theory is UV convergent, we can generate random probability
densities chosen from the prior Eq.~(\ref{prior}) by replacing $\phi$
with its Fourier series and truncating the latter at some sufficiently
high wavenumber $k_c$ [$k_c=1000$ for
Figs.~(\ref{correct_ex}--\ref{incorrect_diff}), and $5000$ for
Figs.~(\ref{scale_select}, \ref{adaptive_D})]. Then Eq.~(\ref{prior})
enforces the amplitude of the $k$'th mode ($k>0$) to be distributed a
priori normally around zero with the standard deviation
\begin{equation}
\sigma_k=\frac{2^{1/2}}{\ell^{\eta-1/2}} 
\left(\frac{L}{2\pi k}\right)^{\eta}.
\label{sigmak}
\end{equation}
Once all these amplitudes are selected, the $k=0$ harmonic is then
set by the normalization condition.

Coded in such a way, the simulations are extremely computationally
intensive because each iteration steps involves an inversion of a
large matrix. Therefore, Monte Carlo averagings given here are only
over 500 runs, fluctuation determinants are calculated according to
Eq.~(\ref{action}), not using numerical path integration, and $Q_{\rm
  cl}=(1/\ell_0)\exp [-\phi_{\rm cl}]$ is always used as an
approximation to $Q_{\rm est}$.

\section{Simulations: Correct prior}

\begin{figure}[t]
  \centerline{\epsfxsize=.9\hsize\epsffile{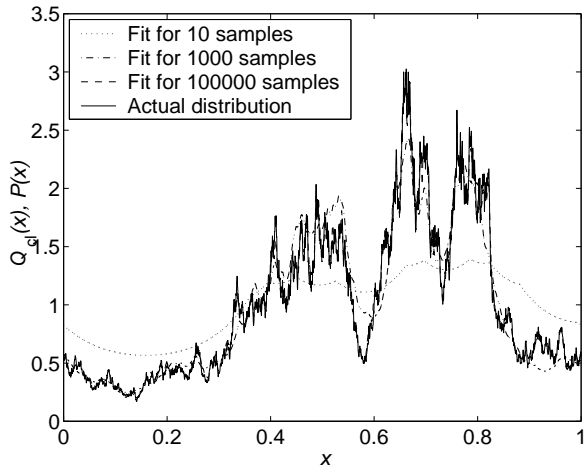}}
\caption{$Q_{\rm cl}$ found for different
  $N$ at $\ell=0.2$.}
\label{correct_ex}
\end{figure}

As an example of the algorithm's performance, Fig.~(\ref{correct_ex})
shows one particular learning run for $\eta =1$ and $\ell=0.2$. We see
that singularities and overfitting are absent even for $N$ as low as
$10$. Moreover, the approach of $Q_{\rm cl}(x)$ to the actual
distribution $P(x)$ is remarkably fast: for $N=10$, they are similar;
for $N=1000$, very close; for $N=100000$, one needs to look carefully
to see the difference between the two.

To quantify this similarity of distributions, we compute the
Kullback--Leibler divergence $D_{\rm KL}(P||Q_{\rm est})$ between the
true distribution $P(x)$ and its estimate $Q_{\rm est}(x)$, and then
average over the realizations of the data points and the true
distribution. As discussed in \cite{bnt}, this learning curve
$\Lambda(N)$ measures the (average) excess cost incurred in coding the
$N+1$'st data point because of the finiteness of the data sample, and
thus can be called the ``universal learning curve''. If the inference
algorithm uses all of the information contained in the data that is
relevant for learning (``predictive information'' \cite{bnt}), then
\cite{bnt,bcs,aida,thesis}
\begin{equation}
\Lambda (N) 
\sim (L/\ell)^{1/2\eta} N^{1/2\eta-1}.
\label{lcurve}
\end{equation}

\begin{figure}[t]
  \centerline{\epsfxsize=.9\hsize\epsffile{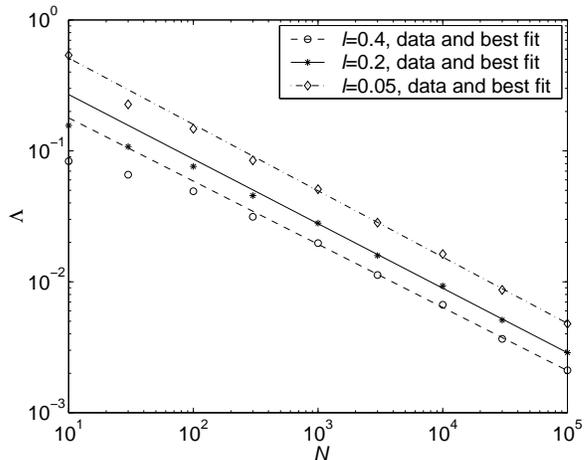}}
\caption{$\Lambda$ as a function of $N$ and $\ell$.
  The best fits are: for $\ell=0.4$, $\Lambda = (0.54 \pm 0.07)
  N^{-0.483 \pm 0.014}$; for $\ell=0.2$, $\Lambda = (0.83 \pm 0.08)
  N^{-0.493 \pm 0.09}$; for $\ell=0.05$, $\Lambda = (1.64 \pm 0.16)
  N^{-0.507 \pm 0.09}$. 
}
\label{correct}
\end{figure}

We test this prediction against the learning curves in the actual
simulations.  For $\eta=1$ and $\ell=0.4,\,0.2,\,0.05$, these are
shown on Fig.~(\ref{correct}).  One sees that the exponents are
extremely close to the expected $1/2$, and the ratios of the
prefactors are within the errors from the predicted scaling $\sim
1/\sqrt{\ell}$.  All of this means that the proposed algorithm for
finding densities not only works, but is at most a constant factor
away from being optimal in using the predictive information of the
sample set.

\section{Simulations: Wrong prior}

\begin{figure}[t]
  \centerline{\epsfxsize=.9\hsize\epsffile{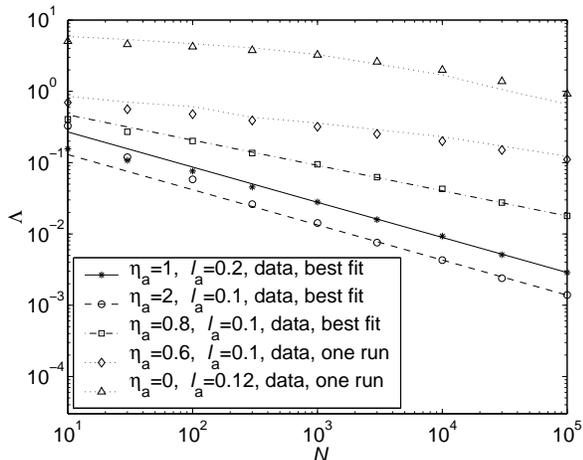}}
\caption{$\Lambda$ as a function of $N$, $\eta_a$ and $\ell_a$.  
  Best fits: for $\eta_a=2$, $\ell_a=0.1$, $\Lambda = (0.40 \pm 0.05)
  N^{-0.493 \pm 0.013}$; for $\eta_a=0.8$, $\ell_a=0.1$, $\Lambda =
  (1.06 \pm 0.08) N^{-0.355\pm 0.008}$.  $\ell=0.2$ for all graphs,
  but the one with $\eta_a=0$, for which $\ell=0.1$.  }
\label{incorrect_diff}
\end{figure}

Next we investigate how one's choice of the prior influences learning.
We first stress that there is no such thing as a {\em wrong prior}. If
one admits a possibility of it being wrong, then it does not encode
all of the a priori knowledge!  It does make sense, however, to ask
what happens if the distribution we are trying to learn is an extreme
outlier in the prior ${\cal P}[\phi]$.  One way to generate such an
example is to choose a typical function from a different prior ${\cal
  P}'[\phi ]$, and this is what we mean by ``learning with a wrong
prior.''
\begin{figure}[t]
  \centerline{ \epsfxsize=.9\hsize\epsffile{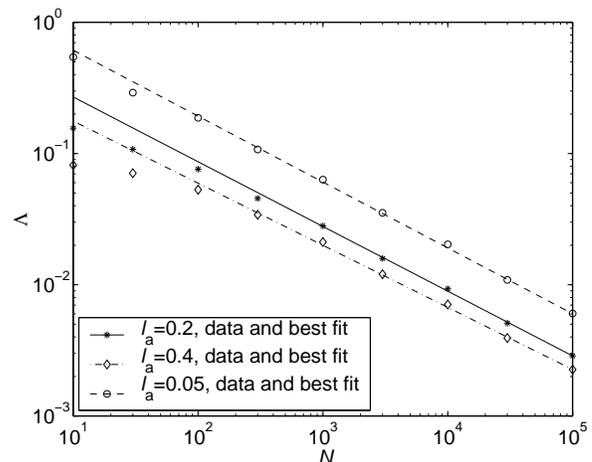}}
\caption{$\Lambda$ as a function of $N$ and $\ell_a$. Best fits are: for
  $\ell_a=0.4$, $\Lambda = (0.56 \pm 0.08) N^{-0.477 \pm 0.015}$; for
  $\ell_a=0.05$, $\Lambda = (1.90 \pm 0.16) N^{-0.502 \pm 0.008}$.
  Learning is always with $\ell=0.2$.
  }
\label{incorrect_same}
\end{figure}
If the prior is wrong in this sense, and learning is described by
Eqs.~(\ref{solution}--\ref{stationary}), then we still expect the
asymptotic behavior, Eq.~(\ref{lcurve}), to hold; only the prefactors
of $\Lambda$ should change, and those must increase since there is an
obvious advantage in having the right prior; we illustrate this in
Figs.~(\ref{incorrect_same}, \ref{incorrect_diff}).

For Fig.~(\ref{incorrect_same}), both ${\cal P}'[\phi ]$ and ${\cal
  P}[\phi ]$ are given by Eq.~(\ref{prior}), but ${\cal P}'$ has the
``actual'' smoothness scale $\ell_a= 0.4,\, 0.05$, and for ${\cal P}$
the ``learning'' smoothness scale is $\ell=0.2$ (we show the case
$\ell_a=\ell=0.2$ again as a reference).  The $\Lambda \sim
1/\sqrt{N}$ behavior is seen unmistakably. The prefactors are a bit
larger (unfortunately, insignificantly) than the corresponding ones
from Fig.~(\ref{correct}), so we may expect that the ``right'' $\ell$,
indeed, provides better learning (see later for a detailed
discussion). 

Further, Fig.~(\ref{incorrect_diff}) illustrates learning when not
only $l$, but also $\eta$ is ``wrong'' in the sense defined above. We
illustrate this for $\eta_a=2,\,0.8,\,0.6,\,0$ (remember that only
$\eta_a>0.5$ removes UV divergences).  Again, the inverse square root
decay of $\Lambda$ should be observed, and this is evident for
$\eta_a=2$. The $\eta_a=0.8,0.6,0$ cases are different: even for $N$
as high as $10^5$ the estimate of the distribution is far from the
target, thus the asymptotic regime is not reached. This is a crucial
observation for our subsequent analysis of the smoothness scale
determination from the data. Remarkably, $\Lambda$ (both averaged and
in the single runs shown) is monotonic, so even in the cases of {\it
  qualitatively} less smooth distributions {\it there still is no
  overfitting}. On the other hand, $\Lambda$ is well above the
asymptote for $\eta=2$ and small $N$, which means that initially too
many details are expected and wrongfully introduced into the estimate,
but then they are almost immediately ($N \sim 300$) eliminated by the
data.

\section{Smoothness scale selection}

Following the argument suggested in \cite{bcs}, we now view ${\cal
  P}[\phi]$, Eq.~(\ref{prior}), as being a part of some wider model
that involves a prior over $\ell$.  The details of the prior are
irrelevant, however, if $S_{\rm eff} (\ell)$, Eq.~(\ref{action}), has
a minimum that steepens as $N$ grows. We explicitly note that this
mechanism is {\em not} tuning of the prior's parameters, but Bayesian
inference at work: $\ell^*$ emerges in a competition between the
kinetic, the data, and the Occam terms to make $S_{\rm eff}$ smaller,
and thus the {\em total} probability of the data is larger. In its
turn, larger probability means, roughly speaking, a shorter total code
length, hence the relation to the MDL paradigm \cite{rissanen}.

The data term, on average, is equal to $ND_{\rm KL}(P||Q_{\rm cl})$,
and, for very regular $P(x)$ (an implicit assumption in \cite{bcs}),
it is small.  Thus only the kinetic and the Occam terms matter, and
$\ell^* \sim N^{1/3}$\cite{bcs}.  For less regular distributions
$P(x)$, this is not true [cf.\ Fig.~(\ref{incorrect_diff})]. For
$\eta=1$, $Q_{\rm cl}(x)$ approximates large-scale features of $P(x)$
very well, but details at scales smaller than $\sim\sqrt{\ell/NL}$ are
averaged out.  If $P(x)$ is taken from the prior, Eq.~(\ref{prior}),
with some $\eta_a$, then these details fall off with the wave number
$k$ as $\sim k^{-\eta_a}$.  Thus the data term is $\sim N^{1.5-\eta_a}
\ell^{\eta_a-0.5}$ and is not necessarily small. For $\eta_a <1.5$
this  dominates the kinetic term and competes with the fluctuations
to set
\begin{equation}
\ell^* \sim N^{(\eta_a -1)/\eta_a},\,\,\,\,\, \eta_a < 1.5\,.
\label{l_dominant}
\end{equation}
There are two remarkable things about Eq.~(\ref{l_dominant}). First,
for $\eta_a=1$, $\ell^*$ stabilizes at some constant value, which we
expect to be equal to $\ell_a$. Second, even for $\eta \neq \eta_a$,
Eqs.~(\ref{lcurve}, \ref{l_dominant}) ensure that $\Lambda$ scales as
$\sim N^{1/2 \eta_a -1}$, which is at worst a constant factor away
from the best scaling, Eq.~(\ref{lcurve}), achievable with the
``right'' prior, $\eta=\eta_a$.  So, by allowing $\ell^*$ to vary with
$N$ we can correctly capture the structure of models that are
qualitatively different from our expectations ($\eta \neq \eta_a$) and
produce estimates of $Q$ that are extremely robust to the choice of
the prior.  To our knowledge, this feature has not been noted before
in a reference to a nonparametric problem.

\begin{figure}[t]
  \centerline{\epsfxsize=.9\hsize\epsffile{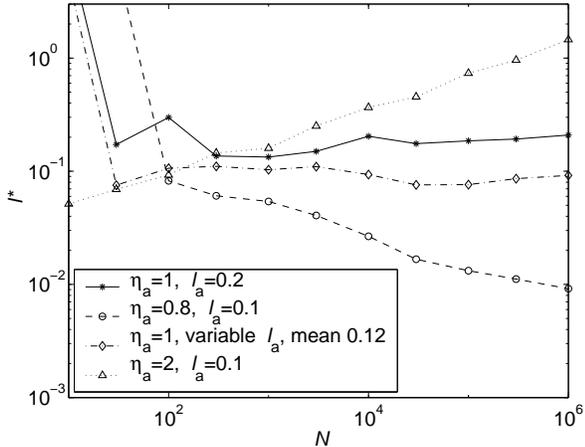}}
\caption{Smoothness scale selection by the data. The lines that go off
  the axis for small $N$ symbolize that $S_{\rm eff}$ monotonically
  decreases as $\ell \to \infty$.} 
\label{scale_select}
\end{figure}

\begin{figure}[b]
  \centerline{\epsfxsize=.9\hsize\epsffile{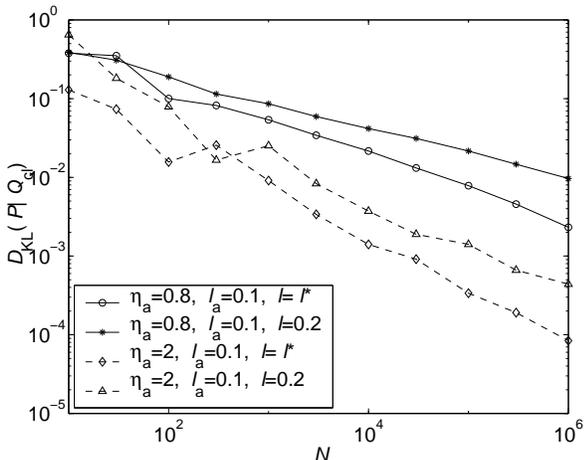}}
\caption{Comparison of learning speed for the same data sets with
  different a priori assumptions.}
\label{adaptive_D}
\end{figure}

We present simulations relevant to these predictions in
Figs.~(\ref{scale_select},~\ref{adaptive_D}). Unlike on the previous
Figures, the results are not averaged due to extreme computational
costs, so all our further claims have to be taken cautiously. On the
other hand, selecting $\ell^*$ in single runs has some practical
advantages: we are able to ensure the best possible learning for any
realization of the data.  Fig.~(\ref{scale_select}) shows single
learning runs for various $\eta_a$ and $\ell_a$. In addition, to keep
the Figure readable, we do not show runs with
$\eta_a=0.6,0.7,1.2,1.5,3$, and $\eta_a \to \infty$, which is a
finitely parameterizable distribution.  All of these display a good
agreement with the predicted scalings: Eq.~(\ref{l_dominant}) for
$\eta_a < 1.5$, and $\ell^*\sim N^{1/3}$ otherwise. Next we calculate
the KL divergence between the target and the estimate at
$\ell=\ell^*$; the average of this divergence over the samples and the
prior is the learning curve [cf.\ Eq.\ (\ref{lcurve})].
For $\eta_a=0.8,2$ we plot the divergences on Fig.~(\ref{adaptive_D})
side by side with their fixed $\ell=0.2$ analogues.  Again, the
predictions clearly are fulfilled.  Note, that for $\eta_a\neq\eta$
there is a {\em qualitative} advantage in using the data induced
smoothness scale.

\section{Parameterization as a wrong prior}

The last four Figures have illustrated some aspects of learning with
``wrong'' priors. However, all of our results may be considered as
belonging to the ``wrong prior'' class. Indeed, the actual probability
distributions we used were not nonparametric continuous functions with
smoothness constraints, but were composed of $k_c$ Fourier modes, thus
had $2k_c$ parameters. For finite parameterization, asymptotic
properties of learning usually do not depend on the priors (cf.\ 
\cite{rissanen,bnt}), and priorless theories can be considered
\cite{vapnik}. In such theories it would take well over $2k_c$ samples
to even start to close down on the actual value of the parameters, and
yet a lot more to get accurate results.  However, using the wrong
continuous parameterization [$\phi(x)$] we were able to obtain good
fits for as low as $1000$ samples [cf.~Fig.~(\ref{correct_ex})] with
the help of the prior Eq.~(\ref{prior}). Moreover, learning happened
continuously and monotonically without huge chaotic jumps of
overfitting that necessarily accompany any brute force parameter
estimation method at low $N$. So, for some cases, a {\em seemingly
  more complex model} is actually {\em easier} to learn!

Thus our claim: when data are scarce and the parameters are abundant,
one gains even by using the regularizing powers of wrong priors. The
priors select some large scale features that are the most important to
learn first and fill in the details as more data become available (see
\cite{bnt} on relation of this to the Structural Risk Minimization
theory). If the global features are dominant (arguably, this is
generic), one actually wins in the learning speed
[cf.~Figs.~(\ref{correct}, \ref{incorrect_same}, \ref{adaptive_D})].
If, however, small scale details are as important, then one at least
is guaranteed to avoid overfitting [cf.~Fig.~(\ref{incorrect_diff})].

One can summarize this in an Occam-like fashion \cite{bnt}: if two
models provide equally good fits to data, {\em a simpler one should
  always be used}. In particular, the predictive information, which
quantifies complexity \cite{bnt}, and of which $\Lambda$ is the
derivative, in a QFT model is $\sim N^{1/2\eta}$, and it is $\sim k_c
\log N$ in the parametric case. So, for $k_c>N^{1/2\eta}$, one should
prefer a ``wrong'' QFT formulation to the correct one. These results are
very much in the spirit of our whole program: not only is the value of
$\ell^*$ selected that simplifies the description of the data, but the
continuous parameterization itself serves the same purpose.

\section{Summary}

The field theoretic approach to density estimation not only
regularizes the learning process but also allows the self-consistent
selection of smoothness criteria through an infinite dimensional
version of the Occam factors. We have shown numerically, and then
explained analytically that this works, even more clearly than was
conjectured: for $\eta_a<1.5$, $\Lambda$ truly becomes a property of
the data, and not of the Bayesian prior!  If we can extend these
results to other $\eta_a$ and combine this work with the
reparameterization invariant formulation of
\cite{periwalshort,periwallong}, this should give a complete theory of
Bayesian learning for one dimensional distributions, and this theory
has no arbitrary parameters.  In addition, if this theory properly
treats the limit $\eta_a \to \infty$, we should be able to see how the
well--studied finite dimensional Occam factors and the MDL principle
arise from a more general nonparametric formulation.

\section*{Acknowledgements}
We thank Vijay Balasubramanian, Curtis Callan, Adrienne Fairhall, Tim
Holy, Jonathan Miller, Vipul Periwal, and Steve Strong for enjoyable
discussions. A preliminary account of this work was presented at the
2000 Conference on Advances in Neural Information Processing Systems
(NIPS--2000) \cite{nips}, and we thank many participants of the
conference for their helpful comments. Work at Princeton was supported
in part by funds from NEC.

\end{document}